\documentclass[twocolumn]{aastex61}

\received{February 26, 2017}
\accepted{March 21, 2017}
\submitjournal{ApJL}

\shorttitle{Dusty winds in AGN}
\shortauthors{S. F. H{\"o}nig \& M. Kishimoto}

\begin{document}

\title{Dusty winds in active galactic nuclei: reconciling observations with models}

\correspondingauthor{Sebastian F. H{\"o}nig}
\email{S.Hoenig@astro.soton.ac.uk}

\author[0000-0002-6353-1111]{Sebastian F. H{\"o}nig}
\affil{Department of Physics \& Astronomy, University of Southampton, Southampton, SO17 1BJ, UK}

\author{Makoto Kishimoto}
\affil{Department of Physics, Kyoto Sangyo University, Motoyama, Kamigamo, Kita-ku, Kyoto, 603-8555, Japan}



\begin{abstract}
This letter presents a revised radiative transfer model for the infrared (IR) emission of active galactic nuclei (AGN). While current models assume that the IR is emitted from a dusty torus in the equatorial plane of the AGN, spatially resolved observations indicate that the majority of the IR emission from $\la$100\,pc in many AGN originates from the polar region, contradicting classical torus models. The new model \textit{CAT3D-WIND} builds upon the suggestion that the dusty gas around the AGN consists of an inflowing disk and an outflowing wind. Here, it is demonstrated that (1) such disk+wind models cover overall a similar parameter range of observed spectral features in the IR as classical clumpy torus models, e.g. the silicate feature strengths and mid-IR spectral slopes, (2) they reproduce the $3-5\,\micron$ bump observed in many type 1 AGN unlike torus models, and (3) they are able to explain polar emission features seen in IR interferometry, even for type 1 AGN at relatively low inclination, as demonstrated for NGC3783. These characteristics make it possible to reconcile radiative transfer models with observations and provide further evidence of a two-component parsec-scaled dusty medium around AGN: the disk gives rise to the $3-5\,\micron$ near-IR component, while the wind produces the mid-IR emission. The model SEDs will be made available for download.
\end{abstract}

\keywords{galaxies: active --- infrared: galaxies --- radiative transfer --- galaxies: individual (NGC3783)}



\section{Introduction} \label{sec:intro}

It was held that the infrared (IR) emission from active galactic nuclei (AGN) originates from a "dusty torus" around the central black hole and accretion disk \citep{Ant93,Net16}. This geometrically thick structure would explain the relative strength of the IR emission as well as the angle-dependent obscuration of the accretion disk and broad line region, leading to the well-established type 1/type 2 dichotomy of unobscured and obscured objects. Radiative transfer models based on such a toroidal geometry have proven successful in reproducing the IR spectral energy distribution (SED) of AGN \citep[e.g.][]{Fri06,Ram09,Alo11}. However, given the milliarcsecond sizes of the IR emission in most AGN, the true geometric distribution of the emission was never scrutinised up until recently.

Advances in IR interferometry made it possible to resolve the near- and mid-IR emission in about three dozen AGN \citep[for a recent overview, see][]{Bur16}. A growing number of nearby Seyfert galaxies have a sufficient interferometric measurements to determine the wavelength- and angular-dependent size of the emission. It has been shown that contrary to the classical torus picture, the bulk of the mid-IR emission originates from the polar region of the AGN rather than the equatorial plane \citep[e.g.][]{Rab09,Hon12,Hon13,Tri14,Lop16}. Detailed studies of the type 1 AGN NGC3783 and the type 2 sources NGC 1068 and the Circinus galaxy imply that the IR emission emerges from at least two distinct parsec-scale structures: (1) a compact, geometrically thin disk in the equatorial region of the AGN, and (2) an extended, elongated polar structure, which is cospatial with the outflow region of the AGN on larger scales \citep{Asm16}. The disk seems to be contributing a relatively larger fraction to shorter wavelengths, while the polar region dominates the overall IR emission energetically, which requires a significant covering fraction. Individual aspects of this behaviour may be explained with radiative transfer effects within a torus; however, the combination of extension, direction and emitted radiation strongly disfavours current torus models, either smooth or clumpy.

One of the hypotheses put forward to explain this structure leans on the idea that matter accreting onto the AGN is subject to radiative and hydrodynamic pressure producing strong outflows \citep[e.g.][]{Kon94,Elv00,Rot12}. In the specific picture outlined in \citet{Hon12}, the disk component represents the inflowing dusty gas. Near the sublimation radius, the radiation pressure from both the AGN and the dust itself causes material to lift up and form a hollow outflow cone. Both outflow and disk contribute to the obscuration. This paper presents the radiative transfer model \textit{CAT3D-WIND} of a clumpy disk and outflow with the geometry and dust composition built upon the hypothesis outlined in \citet{Hon13}. It will be shown that such a configuration reproduces key observational characteristics of AGN, including the overall IR SEDs, the type 1/type 2 dichotomy, the observed disk plus polar emission structure, and the $3-5\,\micron$ bump seen in many unobscured AGN.

\section{CAT3D-WIND: disk + wind model description}

\begin{figure*}
\begin{center}
\epsscale{1.15}
\plottwo{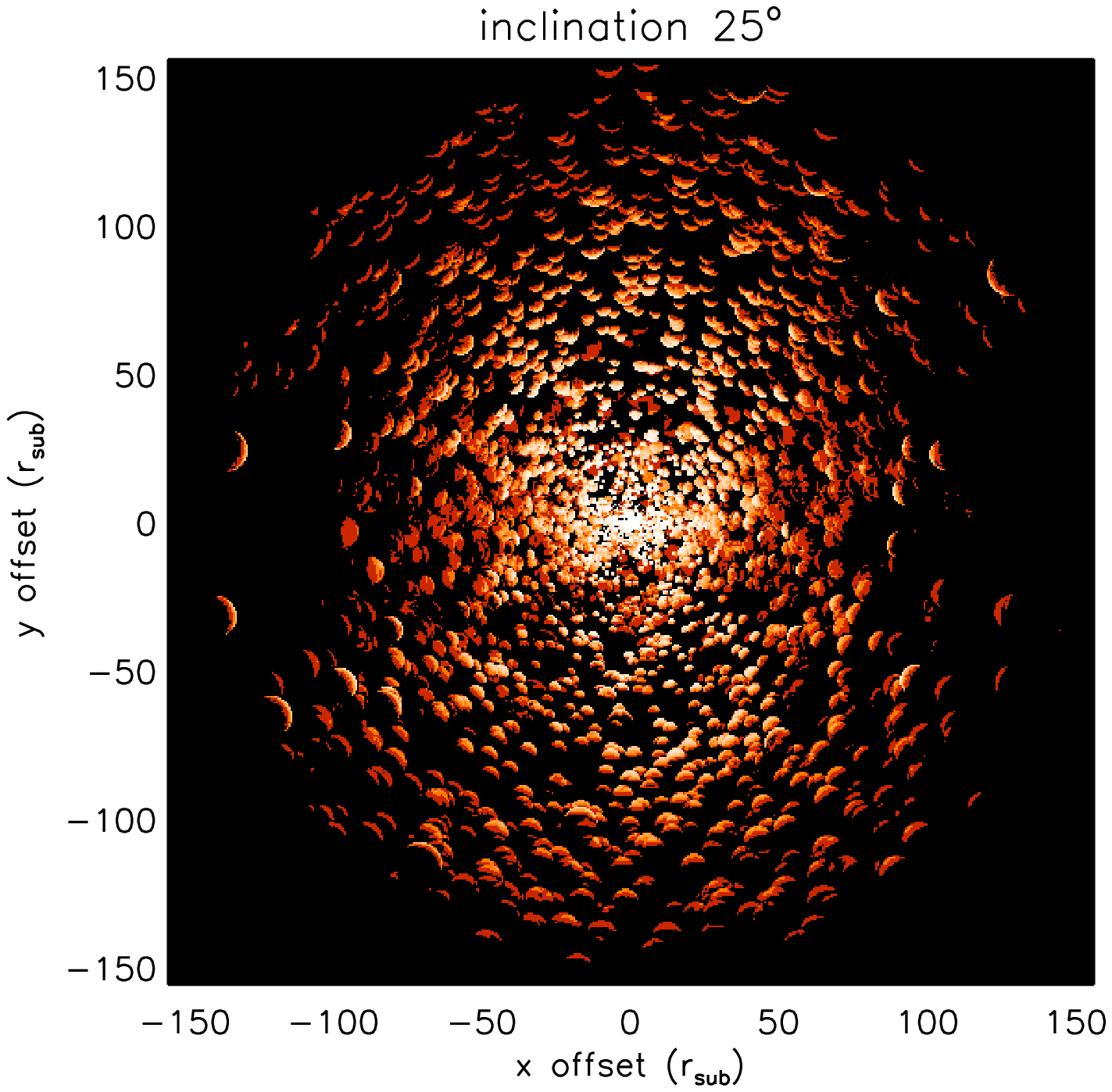}{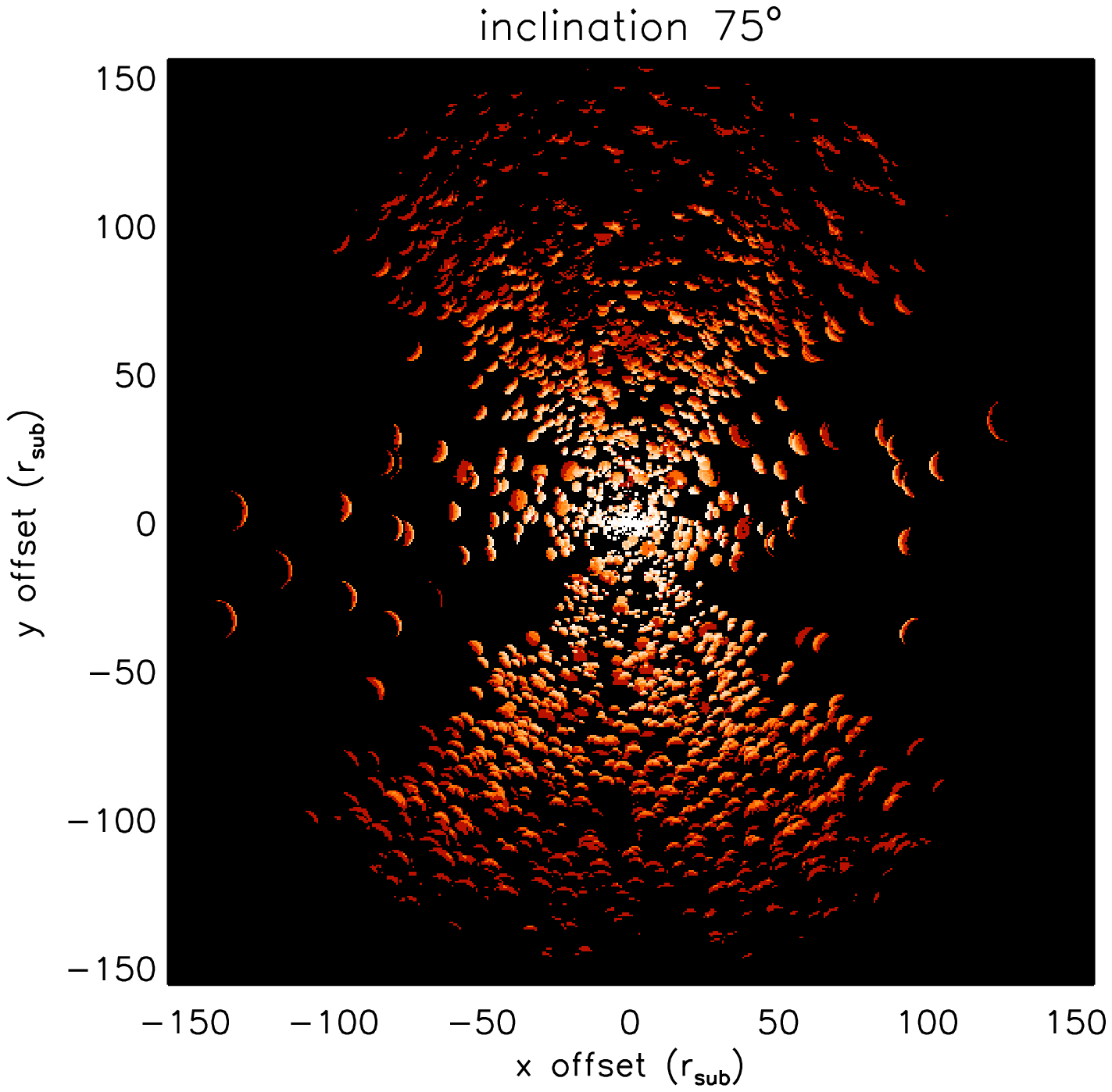}
\end{center}
\caption{Illustrating radiative transfer images of the dust cloud distribution in \textit{CAT3D-WIND} for a compact disk ($a=-3$) and an extended wind ($a_w = -0.5$) with a half-opening angle of $\theta_w=35\degr$. These images represent the histogram-equalised re-emission at $12\,\micron$ for inclinations $i=25\degr$ (view into the cone) and $i=75\degr$ (near edge-on).}\label{fig:illu}
\end{figure*}

The \textit{CAT3D-WIND} model is based on the parametrisation of a clumpy dust torus used in the underlying \textit{CAT3D} model \citep[for details see][]{Hon10b}. It distributes dust clumps according to a radial power law $\propto r^a$, where $a$ is the power-law index and $r$ denotes the distance from the AGN in units of the sublimation radius $r_\mathrm{sub}$. The sublimation radius forms the inner boundary of the dust distribution where the dust becomes too hot to survive. The vertical distribution is characterised by a Gaussian distribution $\propto \exp(-z^2/(2(hr)^2))$ with dimensionless scale height $h$ and vertical distance from the mid-plane $z$ in units of $r_\mathrm{sub}$. The model space is limited from $r_\mathrm{sub}$ as the inner boundary to the outer radius $R_\mathrm{out}$. Please note that $R_\mathrm{out}$ should not be considered a free model parameter, but rather a minimum boundary that has to be chosen to not cause an artificial brightness cut-off in the wavelength range of interest \citep[for a detailed discussion of this issue, see][Sect. 4.1.4]{Hon10b}. Obscuration is defined by the average number of dust clouds $N_0$ along an equatorial line-of-sight. The clouds are characterised by their optical depth $\tau_\mathrm{cl}$ and radius $R_\mathrm{cl}$, but the size of the clouds only controls the total number of clouds and not the typical model SED \citep[we choose a power law of the form $R_\mathrm{cl} = R_\mathrm{cl;0} \cdot r$ for numerical reasons; for further details see][]{Hon10b}. Moreover, as long as the cloud is optically thick in the IR, which most clumpy models assume, the dominating emission from the hot side of the clouds is insensitive to the exact choice of $\tau_\mathrm{cl}$. In summary, these 6 parameters fully define typical clumpy torus models.

Classical torus models do not account for the observed predominance of polar emission in the mid-IR emission of AGN. Thus, a second component is added to the standard model in the form of a polar outflow. This structure is modelled as a hollow cone and characterised by three parameters: (1) the radial distribution of dust clouds in the wind $a_w$, (2) the half-opening angle of the wind $\theta_w$, and (3) its angular width $\sigma_\theta$. Finally, a wind-to-disk ratio $f_{wd}$ defines the ratio between number of clouds along the cone and $N_0$. Note that converting $f_{wd}$ to a mass ratio would require knowledge of $\tau_\mathrm{cl}$ and $R_\mathrm{cl}$. However, these are not constrained by the SEDs. For each set of model parameters, the emission as seen from inclinations of $0\degr-90\degr$ is simulated, in steps of $15\degr$. An illustration of the dust distribution and typical view of a disk+wind is shown in Fig.~\ref{fig:illu}.

The new model makes use of a more physical description of dust sublimation near the AGN than the commonly assumed single sublimation radius. It accounts for the facts that for any given density, the silicate sublimation temperature is lower than the graphite sublimation temperature \citep[e.g.][]{Phi89} and that small grains are hotter for a fixed distance from the AGN. When the dust heats above 1200\,K, silicates are removed from the dust composition, leaving only graphites that can heat up to 1900\,K. In the intermediate stage from 1200\,K leading up to 1900\,K, the smallest graphite grains are removed, so that at the innermost radius where dust can survive only grains with a size between 0.075$\,\micron$ and 1$\,\micron$ are present. In this framework, it is possible to reconcile the observed small near-IR reverberation mapping sizes and near-IR interferometry sizes with dust sublimation physics \citep[e.g.][]{Kis07}. 

One specific aspect of the disk+wind hypothesis put forward in \citet{Hon13} is that the wind is launched near the sublimation zone of the dusty disk. Thus, the chemical composition of the dust in the wind is expected to be very similar to the one seen at about the sublimation radius. Accordingly, the model dust clouds in the polar region are devoid of silicates and small grains. This will also help near edge-on views of the disk+wind structure to display silicate absorption features at $\sim$10\,$\micron$ from the disk despite the direct view to hotter dust in the outflow region. Wind-only models with standard silicate and graphite mixtures predominantly show silicate emission features \citep[e.g.][]{Kea12,Gal15}. This is confirmed with the presented models when using normal ISM dust for the wind clouds. Such models show exclusively silicate emission features for the range of observed SED mid-IR slopes of both type 1 and type 2 AGN (see Sect.~\ref{sec:sed}).

\begin{table}
\begin{center}
\caption{Range of model parameters used to create the parameter grid.}\label{tab:pars}
\begin{tabular}{c l}
\\ \hline
\multicolumn{2}{c}{disk parameters} \\ \hline
$a$ & $\left[-0.5,-1.0,-1.5,-2.0,-2.5,-3.0\right]$ \\
$N_0$ & $\left[5,7.5,10\right]$ \\
$h$  & $\left[0.1,0.2,0.3,0.4,0.5\right]$ \\ \hline
\multicolumn{2}{c}{wind parameters} \\ \hline
$a_w$ & $\left[-0.5,-1.0,-1.5,-2.0,-2.5\right]$ \\
$\theta_w$ & $\left[45\degr,30\degr\right]$ \\
$\sigma_\theta$ & $\left[7.5,10,15\right]$ \\ \hline
\multicolumn{2}{c}{global parameters} \\ \hline
$f_{wd}$ & $\left[0.15,0.3,0.45,0.6,0.75\right]$ \\
               & additionally $\left[1,1.25,1.5,1.75\right]$ for $\theta_w=30\degr$ \\
$R_\mathrm{out}$ $^{(1,2)}$ & $500$ (large graphite) / $151$ (ISM) \\
$R_\mathrm{cl}$ $^{(1)}$ & $0.035 \cdot r$ \\
$\tau_V$ & $50$ \\ \hline
\multicolumn{2}{c}{line of sight} \\ \hline
inclination & $\left[0\degr,15\degr,30\degr,45\degr,60\degr,75\degr,90\degr\right]$ \\ \hline
\end{tabular}
\end{center}
\textit{--- Notes:} For parameter definition see main text. $^{(1)}$ in units of $r_\mathrm{sub}$. $^{(2)}$ The radius where large graphite grains sublimate at 1900\,K is a factor of 3.3 smaller than the classical definition of the sublimation radius. $R_\mathrm{out}$ is set to match the classical \textit{CAT3D} value that does not result in artificial cut-offs in the mid-IR \citep{Hon10b}.
\end{table}

\section{Results and discussion}

The parameter space of the model has been explored in the range of parameters as listed in Table~\ref{tab:pars}. A total of 132,300 model SEDs have been simulated. It should be noted that the range of $f_{wd}$ is larger for $\theta_w = 30\degr$ than for $45\degr$. In general, the covered surface for a cone with half-opening angle closer to the polar axis is smaller than for a cone with a larger half-opening angle. To counter this effect, the effectively filled area in the cone for the models with smaller half-opening angle has been increased.

\subsection{SED shapes and silicate features}\label{sec:sed}

\begin{figure*}
\begin{center}
\epsscale{1.15}
\plottwo{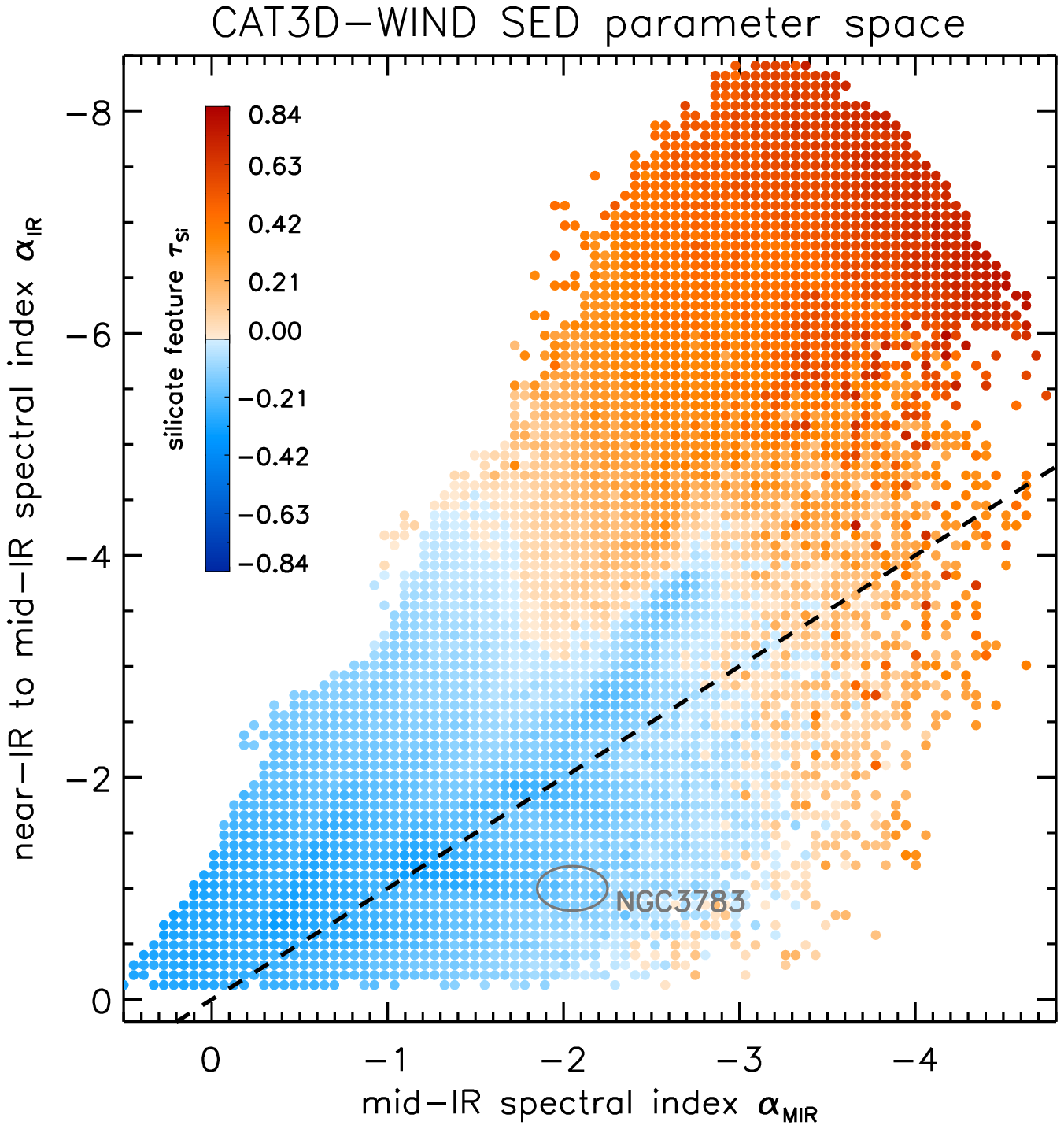}{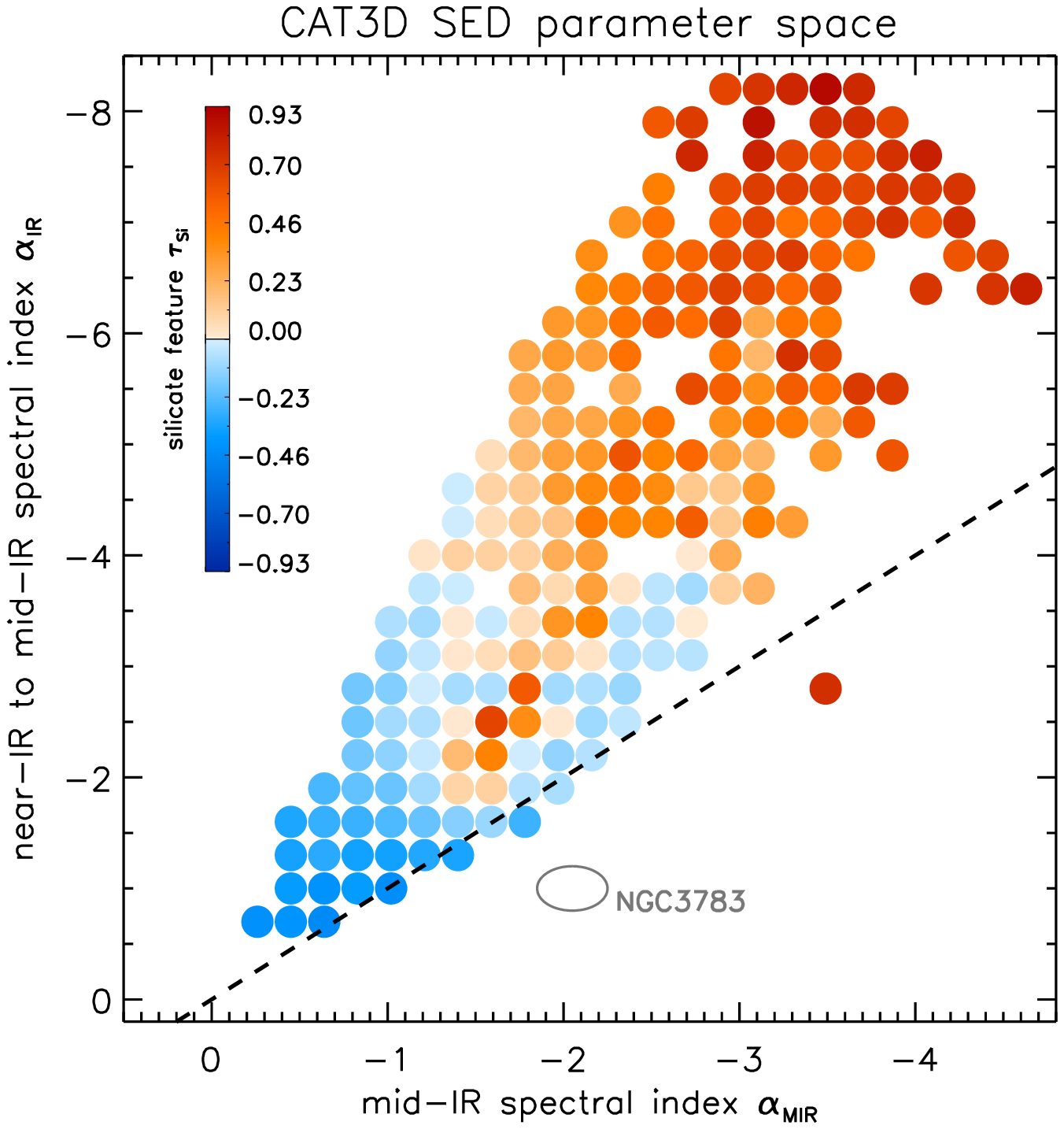}
\end{center}
\caption{Observed SED parameter space covered by the grid of \textit{CAT3D-WIND} models (left) and the classical clumpy torus models of \textit{CAT3D} (right). Each coloured circle represents one or more model SEDs from the respective model grid. While the position in the grid indicates the spectral indices in the mid-IR, $\alpha_\mathrm{MIR}$, and near-to-mid-IR, $\alpha_\mathrm{IR}$ covered by the models, the colour represents the strength of the silicate feature, $\tau_\mathrm{Si}$, with negative numbers (=blue colours) noting an emission feature and positive numbers (=red colours) an absorption feature. The dashed black line marks equal spectral indices. Below this line, the near-IR emission is bluer than the mid-IR, which results in the presence of two distinct emission bumps. The grey contour denotes the model SEDs that are consistent with observations of NGC3783.}\label{fig:seds}
\end{figure*}

To obtain a swift overview of the typical SED shapes of all the models, each model SED has been broken down into three parameters: (1) the spectral index $\alpha_\mathrm{MIR}$ of the mid-IR emission, expressed by the flux ratio between $14\,\micron$ and $8\,\micron$, (2) the spectral index $\alpha_\mathrm{IR}$ from the near-IR to the mid-IR, expressed by the flux ratio between $6\,\micron$ and $3\,\micron$, and (3) the silicate feature optical depth $\tau_\mathrm{Si}$. Here, the spectral index is defined as
\begin{equation}
\alpha = -\frac{\log\left(F_\nu(\lambda_j)/F_\nu(\lambda_i)\right)}{\log\left(\lambda_j/\lambda_i\right)}
\end{equation}
where $F_\nu(\lambda)$ refers to the flux at wavelength $\lambda$ and $\lambda_j > \lambda_i$. Fig.~\ref{fig:seds} shows an overview of the range of SED shapes obtained by the \textit{CAT3D-WIND} model based on the parameter grid in Table~\ref{tab:pars}. In addition, a corresponding grid for the classical clumpy torus based on \textit{CAT3D} is shown for comparison by setting $f_{wd}=0$. Greater spectral indices relate to bluer SEDs. For $\tau_\mathrm{Si}$, blue colours indicate silicate emission features and red colours silicate absorption features. The range of $\alpha_\mathrm{MIR}$ has been studied in the past for a variety of AGN classes \citep[e.g.][]{Hao07,Hon10a,Alo16,Gar17}. In general, Seyfert galaxies and quasars fall in the range of $-0.5 > \alpha_\mathrm{MIR} > -3$, with type 1 AGN showing slightly bluer SEDs than type 2 AGN. This range of observed parameters is well covered by both model grids, illustrating the general viability of the disk+wind models as well as the classical torus models.

One interesting aspect of the covered model space is the relation of $\alpha_\mathrm{MIR}$ and $\alpha_\mathrm{IR}$. Fig.~\ref{fig:seds} shows a black dashed line that marks equal spectral index in the near- and mid-IR. Generally, classic clumpy torus models cover the parameter space above this line, where the near-IR rises relatively steeply towards the mid-IR \citep[see][Figs. 5 \& 6]{Hon10a}. Then, the SEDs flatten in the mid-IR and turn over longward of $\sim20-30\,\micron$. Such models struggle to reproduce the $3-5\,\micron$ bump that is seen in several type 1 AGN, including quasars \citep[e.g.][]{Ede86,Kis11b,Mor12}. Those SEDs are characterised by a first, local emission peak in the near-IR regime of the SED before they become steeper again towards the global $20-30\,\micron$ peak. SEDs with a $3-5\,\micron$ bump are found below the dashed line where the near-to-mid-IR spectral index $\alpha_\mathrm{IR}$ is bluer than the mid-IR spectral index $\alpha_\mathrm{IR}$. Indeed, the disk+wind models of \textit{CAT3D-WIND} naturally produce such SEDs for a significant number of parameter combinations. Most notably, the $3-5\,\micron$ bump emerge from models where the disk has a relatively steep dust cloud distribution of $-2 > a > -3$ while the distribution of clouds in the wind is much shallower with $-0.5 > a > -1.5$. In these cases, the near-IR emission is dominated by the hot dust in the disk and the wind primarily contributes at longer wavelengths, which is the scenario implied by interferometry observations of NGC3783 \citep{Hon13}.

The silicate emission features in most of the \textit{CAT3D-WIND} model SEDs are relatively shallow while the absorption features can be deeper. Qualitatively, this is consistent with observational constraints for both space-based and ground-based AGN samples \citep[e.g.][]{Hao07,Hon10a,Alo16}. The distribution of observed feature strengths peak in the region of about $0.1 \la \tau_\mathrm{Si} \la -0.4 $ for quasars, $0.3 \la \tau_\mathrm{Si} \la -0.3 $ for Seyfert 1 galaxies, and $1 \la \tau_\mathrm{Si} \la -0.1 $ for Seyfert 2 AGN. The model space of the presented parameter grid covers these values.

In summary, the model SEDs of the disk+wind model cover the observed SED shapes and silicate feature depths quite well, very similar to classical clumpy torus models. In addition and beyond the capabilities of the torus models, the disk+wind model is able to reproduce the observed $3-5\,\micron$ bump in type 1 AGN from a compact disk, while preserving the mid-IR bump produced by the wind. This makes it superior to classical clumpy torus models on the criterium of reproducing SEDs alone.

\subsection{Polar elongations in type 1 AGN: NGC3783 as a case study}\label{sec:ngc3783}

\begin{figure}
\begin{center}
\epsscale{1.2}
\plotone{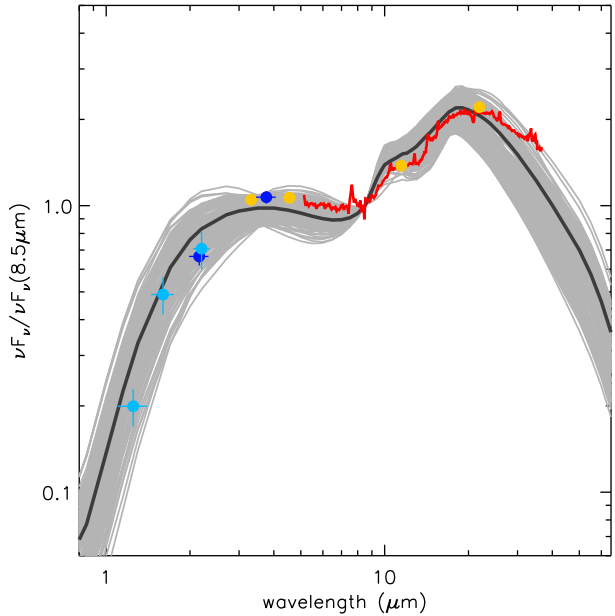}
\end{center}
\caption{Comparison between the observed and model IR SEDs of NGC3783. The observations are collected from 2MASS \citep[light blue, ][]{Wei12}, ISAAC (dark blue), Spitzer IRS (red line) and IRAC \citep[orange, ][]{Hon13}. The light-grey band shows the range of model SEDs consistent with the observed IR and mid-IR slopes of NGC3783. In addition, the dark grey line indicates the model SED for the model used to reproduce the interferometric observations (see Fig.~\ref{fig:ngc3783_sed}).}\label{fig:ngc3783_sed}
\end{figure}

Type 1 AGN NGC3783 displays a pronounced $3-5\,\micron$ bump (see Fig.~\ref{fig:ngc3783_sed}). Decomposing the SED of NGC3783 similarly to the \textit{CAT3D-WIND} model SEDs results in $\alpha_\mathrm{MIR} = -2.05\pm0.20$, $\alpha_\mathrm{IR} = -1.0\pm0.2$, and $\tau_\mathrm{Si} = -0.07 \pm 0.06$. This puts NGC3783 in a range that is only covered by the disk+wind models, not the classical clumpy torus models (see grey contours in Fig.~\ref{fig:seds} left \& right). In Fig.~\ref{fig:ngc3783_sed} we plot the range of model SEDs that fall within the uncertainty interval of these parameters as a grey band. The majority of these models are characterised by a compact, relatively flat disk ($a \sim -2.0\ldots-3$; $h \la 0.3$) and an extended wind ($a_w \ga -1$). Inclinations range from $0\degr \le i \le 60\degr$ indicating a moderately inclined AGN, as can be expected for a type 1. The SED-only models are inconclusive with regards to the half-opening angle of the outflow.

The major motivation for the disk+wind model stems from IR interferometry of nearby AGN. Therefore, viability of the new models rests on the ability to not only reproduce the SEDs but also the interferometry measurements. In general, it can be expected that a type 2 AGN seen edge on will indeed show elongation towards the polar region in these models. On the other hand, radiative transfer effects might also produce some polar emission in obscured AGN in classical torus models \citep[e.g.][]{Sch05,Sch08}. The more challenging objects are moderately-inclined type 1 AGN where classical models would produce an elongated, disk-like structure with orientation along the disk plane.

NGC3783 can be considered a prototypical inclined type 1 AGN with dominating polar IR emission. It has been extensively covered with near- and mid-IR interferometry. \citet{Hon13} report simple modelling of 41 VLTI MIDI observations in the mid-IR and 6 VLTI AMBER measurements in the near-IR \citep[AMBER data originally from][]{Wei12}. The IR SED shows the characteristic $3-5\,\micron$ bump with an estimated hot-dust covering factor of about 30\%. The near-to-mid-IR photometric and interferometric data is best reproduced by two IR emitting components, characterised by two perpendicular elongated structures, with the polar-oriented dominating in the mid-IR. The mid-IR emission is very elongated with a major-to-minor axis ratio of $\sim$3:1, which is challenging to achieve in classical torus models for an unobscured type 1 AGN at moderate inclinations. 

\begin{figure}
\begin{center}
\epsscale{1.2}
\plotone{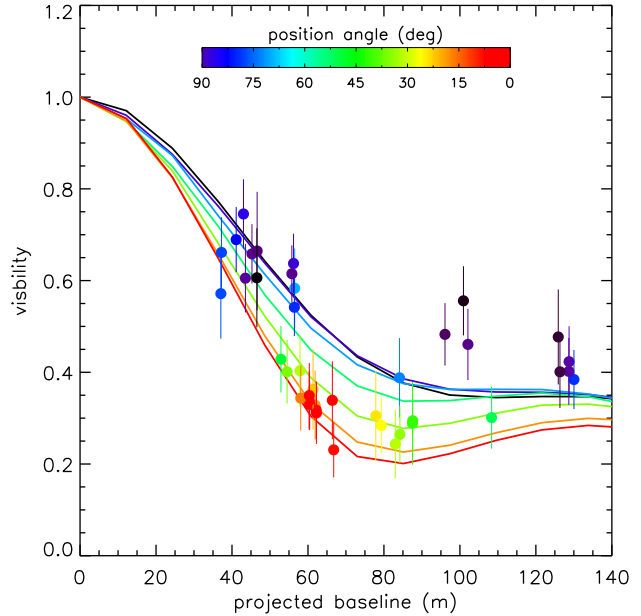}
\end{center}
\caption{Comparison between observed and model $12\,\micron$ interferometry of NGC3783. The different colours note the position angle for each visibility point with $0\degr$ (red) along the polar axis and $90\degr$ (violet) parallel to the equatorial plane. Model parameters are given in the text. The corresponding SED is highlighted in Fig.~\ref{fig:ngc3783_sed}.}\label{fig:ngc3783_intf}
\end{figure}

The best SED model fits (see grey contour in Fig.~\ref{fig:seds}, right) were selected to simulate multi-wavelength images, which is computationally expensive. Interferometric visibilities were obtained for the model images at $12\,\micron$ and compared to the corresponding interferometric visibilities of NGC3783 \citep[see][Fig. 7]{Hon13}. It is immediately found that the strong elongation seen in NGC3783 disfavour low inclinations at $0\degr-15\degr$, as expected. The strongest elongations are seen either at higher inclination, where the cone is seen from the side, or when the line-of-sight into the cone is close to its edge. Stronger polar elongations are found for $\theta_w = 30\degr$ than for $\theta_w = 45\degr$.

Fig.~\ref{fig:ngc3783_intf} shows model visibilities for several position angles for a representative model with $a=-3$, $a_w = -0.5$, $N_0=10$, $\theta_w = 38\degr$, $\sigma_\theta = 10\degr$ and $f_{wd}=1.2$ seen under an inclination of $i=30\degr$. This orientation and cone shape implied by this model are in between the suggestions of \citet{Mul11} and \citet{Fis13}, who infer $i=60\degr/\theta_w = 30\degr$ and $i=15\degr/\theta_w=45\degr$, respectively. While \citet{Fis13} argue that kinematics disfavour the high-inclination solution, the strong elongation seen in IR interferometry might indicate a higher inclination. On the other hand, it is shown here that a configuration where the line-of-sight is close to the cone edge produces strong elongations as well.

\subsection{Anisotropy of the mid-IR emission}

An optically thick dusty torus causes significant anisotropy of the emerging IR emission depending on inclination: More IR radiation escapes towards low inclinations than towards highly inclined line-of-sights. However, when taking the intrinsic X-ray emission as an isotropic tracer for the AGN luminosity, it has been shown that the mid-IR emission is surprisingly isotropic \citep[e.g.][]{Lut04,Gan09,Asm15}, unlike expected from smooth torus models. While this small scatter in the mid-IR-to-X-ray luminosity correlation can be partly explained in the framework of clumpy torus models, the range of torus parameters leading to low anisotropy is very restricted, which would imply a physical mechanism that restricts those parameters \citep[e.g.][]{Hon11a}.

In the disk+wind picture, the hollow cone is visible to the observer at any inclination, naturally leading to low anisotropy of the mid-IR emission. Quantitatively, the peak of the luminosity-normalised flux distributions at $12\,\micron$ shifts by about 0.4\,dex from $0\degr$ to $90\degr$. When considering typical inclinations of $30\degr$ for type 1 AGN and $75\degr$ for type 2s, the anisotropy is $0.24^{+0.11}_{-0.16}$\,dex. This is consistent with estimates from the relation between X-ray and mid-IR luminosities in AGN, finding $<0.3$\,dex \citep{Asm15} and samples isotropically selected in the radio at $\sim0.22$\,dex \citep{Hon11a}.

\section{Summary and Conclusions}

In this letter, a new radiative transfer model \textit{CAT3D-WIND} is presented to reconcile spatially resolved observations of AGN in the IR with model predictions. The model consists of a geometrically thin disk of optically-thick dust clumps and a hollow cone of putatively outflowing dust clouds. It is phenomenologically motivated by recent results in IR interferometry that hint towards such a two-component structure. In this picture, dust clouds are accreted in the plane of the disk. As dust partially sublimates near the sublimation radius (leading to a dearth of silicates), some dust clouds are lifted up by radiation pressure and flow out into the polar region. Based on a parametrisation of this picture, the following has been found:
\begin{itemize}
\item The new disk+wind models cover a similar parameter space in terms of mid-IR spectral slope and strength of the silicate feature as classical clumpy torus models.
\item It is found that the new models are able to explain the $3-5\,\micron$ SED bump seen in many type 1 AGN. This local emission peak appears from the disk when it has a more compact distribution of dust clouds than the wind.
\item Many of the disk+wind models do indeed show polar elongation for inclinations typically associated with both type 1 and type 2 AGN. Indeed, a model has been found to simultaneously reproduce the IR SED and IR interferometry of the type 1 AGN NGC3783.
\item Exposure of the cone to the observer at all inclinations naturally results in a low degree of anisotropy of the mid-IR emission with respect to the inclination or viewing angle. This is consistent with the observed low scatter of the mid-IR/X-ray luminosity correlation.
\end{itemize}
The presented models are radiative transfer models of an empirical hypothesis. However, the physical mechanisms to drive the wind are not clear yet. It is possible that a combination of radiation pressure on the dust and neutral gas, both from the AGN radiation and from the IR-emitting medium itself, will play a role. Furthermore, hydrodynamic pressure close to the sublimation radius may help in launching such a wind. Work on radiative-hydrodynamical simulations will be needed to find a physical foundation for the observed polar structure \citep[e.g.][]{Dor12,Wad12,Wad15,Dor16,Cha16}. The SEDs presented in this paper will be made available for download at \url{http://cat3d.sungrazer.org}.

\acknowledgments

We thank the anonymous referee for helpful comments that improved the manuscript. SFH acknowledges support for this work from the European Research Council Horizon 2020 grant DUST-IN-THE-WIND (677117). Part of the numerical computations were done on the Sciama High Performance Compute (HPC) cluster which is supported by the ICG, SEPNet and the University of Portsmouth. Special thanks go to D. Asmus, P. Gandhi, C. Ricci, M. Stalevski, and K. Tristram for helpful discussions.

%


\software{CAT3D \citep{Hon10b}
          }

\end{document}